# High performance photonic reservoir computer based on a coherently driven passive cavity

Quentin Vinckier,[1,*] François Duport,[1] Anteo Smerieri,[1] Kristof Vandoorne,[2] Peter Bienstman,[2] Marc Haelterman,[1] and Serge Massar[3]

[1]Service OPERA-Photonique, CP 194/5, Université Libre de Bruxelles (U.L.B.), Avenue Adolphe Buyl 87, 1050 Bruxelles, Belgique

[2]Photonics Research Group, Dept. of Information Technology, Ghent University – IMEC, Sint-Pietersnieuwstraat 41, 9000 Gent, Belgium

[3]Laboratoire d'Information Quantique, CP 225, Université Libre de Bruxelles (U.L.B.), Boulevard du Triomphe, 1050 Bruxelles, Belgique

quentin.vinckier@ulb.ac.be

**ABSTRACT**

**Reservoir computing is a recent bio-inspired approach for processing time-dependent signals. It has enabled a breakthrough in analog information processing, with several experiments, both electronic and optical, demonstrating state-of-the-art performances for hard tasks such as speech recognition, time series prediction and nonlinear channel equalization. A proof-of-principle experiment using a linear optical circuit on a photonic chip to process digital signals was recently reported. Here we present a photonic implementation of a reservoir computer based on a coherently driven passive fiber cavity processing analog signals. Our experiment has error rate as low or lower than previous experiments on a wide variety of tasks, and also has lower power consumption. Furthermore, the analytical model describing our experiment is also of interest, as it constitutes a very simple high performance reservoir computer algorithm. The present experiment, given its good performances, low energy consumption and conceptual simplicity, confirms the great potential of photonic reservoir computing for information processing applications ranging from artificial intelligence to telecommunications.**

## 1. Introduction

Reservoir computing is a recent bio-inspired method for processing information [1-3] that uses a recurrent dynamical system called "reservoir" to process time-dependent signals. The internal variables of the dynamical system, also called "reservoir states", provide a nonlinear mapping of the input into a high dimensional space. The time-dependent output of the reservoir is then given by a linear combination of the internal variables. The readout weights used to compute this linear combination are optimized so as to minimize the mean square error between the target and the output signal, leading to a simple and easy training process. Moreover, some global parameters can be tuned to get the best performance, depending on the reservoir architecture and on the task. The simplicity and the flexibility of such systems make them powerful to solve a very large range of different tasks, such as speech recognition [4], nonlinear channel equalization [3,5-7], detection of epileptic seizures [8], robot control [9], time series prediction [1,3,7,10], financial forecasting, handwriting recognition, etc..., see [10,11] for recent reviews.

Recently, experimental implementations of reservoir computing have provided a breakthrough in analog information processing, and in particular in optical information processing. Several experiments [12-22] (see also the simulations in [23]) report analog information processing, often with error rate comparable to the best digital algorithms. These works all use as nonlinear dynamical system a delay line with a single nonlinear node.

Integrated optical RCs (Reservoir Computers) have also been investigated in simulation [24-28]. Very recently a proof-of-principle experiment that carried out simple processing of digital signals was demonstrated [29]. This experiment uses a linear optical network made from interconnected delay lines and works with coherent light, thereby encoding information in both the amplitude and the phase of the electromagnetic field. This can improve performance [24,25] compared to systems that use only the light intensity. The necessary nonlinearity was provided by the readout photodiodes that, since they produce a current proportional to the intensity of the light, provide a quadratic nonlinearity. The use of a nonlinearity in the readout layer of reservoir computers has been previously investigated theoretically [5,6].

Here we combine the advantages of the approach based on a delay dynamical system, and of the approach based on a linear optical circuit with a quadratic nonlinearity provided by the readout photodiode. We present an experimental implementation of a photonic reservoir computer based on a coherently driven passive fiber cavity that processes analog signals. Our system exhibits several advantages compared to previous experiments. First, it is much more flexible than the architecture presented in [29], in which many parameters, such as the number of internal variables or the strength of the internal connections, are fixed by the hardware and cannot be tuned to improve performance on specific tasks. In addition, we are able to process analog signals experimentally (contrary to the experiment described in [29] which only processes digital signals, due to the limitations of very high-speed electronics). Nevertheless, the reservoir presented in [29] is able to process the information faster thanks to the integration on chip. Second, our reservoir is a passive optical cavity, with very low intra-cavity losses. We therefore do not need any amplifier or active element in the cavity as in previous experiments based on a delay line [12-22,30] (for instance in [17], a passive nonlinearity was used, but because of high intra-cavity losses an amplifier was placed inside the cavity). This absence of active elements in the cavity removes a major source of noise, and therefore improves performance. It also decreases the energy consumption of the reservoir. In fact the total optical power injected in the reservoir layer (0.57mW peak power at the entrance of the cavity) is to our knowledge the lowest used so far.

Most importantly, in all tasks on which we have tested it, our experimental reservoir computer has error rate as low or lower than all previous experiments [12-17,21,22,29,31] that studied the same tasks, and the same also holds true for many results obtained previously in simulation or using digital algorithms [7,12,22,24,29,32]. Note that the experiment can also be translated into a digital algorithm. This algorithm is of interest by itself as it constitutes a very simple high performance reservoir computer algorithm.

## 2. Operation principle

In the experiment reported here, the input signal is coded in the amplitude $A_{in}(t)$ of the electromagnetic field. This signal is then sent to the reservoir which consists of a passive fiber cavity (see Fig. 1). The evolution equation of the amplitude $A(t)$ inside the cavity is given by:

$$A(t) = \alpha \exp(j\Delta\varphi)A(t - T) + \beta A_{in}(t), \quad (1)$$

where $T$ is the roundtrip time, $\alpha$ the feedback gain that is tuned through an intracavity optical attenuator, $\beta$ the transmission coefficient of the input fiber coupler, $j=(-1)^{1/2}$, and $\Delta\varphi$ is the phase detuning of the cavity.

As explained in detail in [12,13], successful use of the delay dynamical system architecture as a reservoir computer depends crucially on the method used to encode the input signal. Here we use the method introduced in [13]. The input $u(t)$ is held constant for duration $T'$ using a sample and hold procedure, then this input is multiplied by a mask function $m(t)$ which is periodic with period $T'$. The interconnections between the internal variables are obtained by desynchronizing the masked input $m(t)u(t)$ and the cavity so $T'\neq T$. In the present case, it is further essential for some tasks to add a bias $A_0$ to the masked input. The input signal amplitude is therefore equal to $A_{in}(t)=m(t)u(t)+A_0$. The mask function $m(t)$ is taken to be a step function, constant over intervals of duration $\theta=T'/N$ where $N$ is the number of internal variables of the reservoir. The desynchronisation is defined by $T'=TN/(N+k)$ where $1\leq k<N$. All our simulation and experimental results were obtained with $k=1$, i.e. each internal variable is thus coupled with its nearest neighbor. In this way, we get the richest dynamics possible for this unsynchronized regime (e.g., if $k$ and $N$ have a common divisor, than there would be several independent reservoirs, each with a fraction of the number of neurons. The dynamics would then be less rich, and the performance worse).

In order to discuss in more detail the operation principle of the reservoir, it is useful to introduce a discretized time $t(n,i)$ corresponding to the middle of each interval of duration $\theta$: $t(n,i)=nT'+(i+1/2)\theta$ where $n\in\mathbb{Z}$ and $i\in[0,N\text{-}1]$. With this notation, the sample and hold input $u(n)$ is only a function of $n$, the mask $m(i)$ is only a function of $i$. We denote $x_i(n)=A(t(n,i))$ the state of the internal variables. The continuous time evolution equation (1) can thus be approximated by the discrete time evolution equations:

$$\begin{cases} x_i(n) = \alpha x_{i-k}(n-1)\exp(j\Delta\varphi) + \beta[m_i u(n) + A_0] & k \leq i < N \\ x_i(n) = \alpha x_{N+i-k}(n-2)\exp(j\Delta\varphi) + \beta[m_i u(n) + A_0] & 0 \leq i < k \end{cases},$$
(2)

with $1\leq k<N$. Note that when $\Delta\varphi\neq 0$, $x_i$ are complex values and the dynamics of the system is richer. As discussed in the next section, experimentally it is often convenient to carry out a nonlinear pre-processing of the input. This small change does not modify the performance of the reservoir, except for the evaluation of the memory capacities for which the results depend on this input signal preprocessing.

All the reservoir states denoted $x_i(n)=A(t(n,i))$ are recovered by a photodiode which performs a quadratic transformation on $A(t)$, since the photodiode output is proportional to $|x_i(n)|^2$. The output $y(n)$ of the reservoir is taken to be a linear combination of the $|x_i(n)|^2$:

$$y(n) = \sum_{i=0}^{N-1} W_i |x_i(n)|^2. \quad (3)$$

Note that eqs. (2) and (3) are the numerical model we will refer to later in the text.

In the operation of a reservoir computer we distinguish two phases. In the training phase we send inputs for which the target output $y^*(n)$ is known, we record the values of $|x_i(n)|^2$ and use them to compute the readout weights $W_i$ by minimizing the mean square error $\langle[y^*(n)\text{-}y(n)]^2\rangle_n$ using Tikhonov regularization [13] (also called ridge regression). In the test phase, the readout weights $W_i$ are kept fixed, the output signal $y(n)$ is computed using equation (3), and compared to the target output $y^*(n)$. The performance of the reservoir computer can be optimized by adjusting the values of the parameters $\alpha$, $\Delta\varphi$, $A_0$, and the amplitude of $m_i$ (equivalent to changing $\beta$, which in the experiment is fixed by the coupling ratio of the cavity input coupler and cannot be modified). The input mask $m_i$, randomly chosen from a uniform distribution, is kept fixed.

As we show in the next sections, experiment and numerical simulations based on eqs. (2) and (3) perform very well on a number of benchmark tasks. These good performances can be explained by the following points.

First, using complex variables (when $\Delta\varphi\neq 0$) makes the internal dynamics richer. For some tasks this is crucial, and contributes importantly to the improvement of performance (a fact already pointed out in [24,25,29]). Indeed using complex variables doubles the number of internal variables (2$N$), even if we can only use $N$ readout weights $W_i$ to compute the output $y$.

Then, another contributing fact is that there is very little noise in our experiment. This is confirmed by the total memory capacity being close to its maximum value (see "Results" section). This weak noise level is due to the absence of active (and therefore noise inducing) elements in the reservoir.

Finally, a third favorable factor has more to do with the tasks investigated than with the reservoir itself. Indeed a reservoir computer can be seen as a map from the input time series $u(n)$ to the variables used to compute the output (here $|x_i(n)|^2$). The output of the reservoir is the projection of the desired output $y^*(n)$ onto the space spanned by the $|x_i(n)|^2$ (see [33] for a detailed discussion). In the present case the reservoir states $x_i(n)$ are given by a linear combination of the previous inputs $u(n-l)$ ($l=0,1,2,...$) with exponentially decaying dependence on $l$ (the fading memory property essential for good reservoir performance). The variables $|x_i(n)|^2$ used to compute the output are therefore given by the sum of a constant, a linear, and a quadratic fading memory function of previous inputs (note the role of the bias $A_0$ that ensures that $|x_i(n)|^2$ contain linear terms). The fact that a reservoir based on eqs. (2) and (3) performs well on a task therefore means that for this task, the desired output $y^*(n)$ can be well approximated by quadratic functions of previous inputs. The higher order non-linearities present in many reservoir algorithms are therefore not necessary for these tasks, and may even be detrimental. There are of course some tasks that require non-linearities of higher order than quadratic, and for these tasks our architecture would not give good performance. However by cascading reservoirs of the type reported here, non-linearities of arbitrarily high order could be obtained, so this is not a fundamental problem.

## 3. Experimental implementation

Our fiber optics experiment is depicted in Fig. 1. The optical signal is generated by a CW (Continuous Wave) laser at 1550nm with a coherence time much greater than the inverse linewidth of the cavity, with output power adjusted between 11 and 41mW. Note that below we always quote the optical power at the entrance of the cavity, as this is the experimentally relevant quantity, and the setup was not optimized to minimize the losses between the laser and the cavity.

The input signal $A_{in}(t)$ is encoded by modulating the amplitude of the laser signal with a Lithium Niobate Mach-Zehnder (M-Z) interferometer in push-pull configuration driven by an Arbitrary Waveform Generator (AWG). Consider first that no bias $A_0$ is needed. As the M-Z transfer function is sinusoidal, we first tried to precompensate the voltage signal $V$ generated by the AWG, so that $A_{in}(t)$ is proportional to $m(t)u(t)$. Then we removed the precompensation of the masked input signal, so that the optical M-Z output signal $A_{in}(t)$ is proportional to $\sin\{V(t)[\pi/(2V_\pi)]\}$ with $V(t)= \gamma m(t)u(t) \in [-\gamma V_\pi, \gamma V_\pi]$ ($V_\pi$ being the characteristic voltage of the M-Z modulator and $\gamma \in [0,1]$ is an adjustable parameter). We have checked that both codings give the same performance of the reservoir, except for the evaluation of the memory capacities for which the results depend on this input signal preprocessing.

For some tasks, it is necessary to bias the input. Experimentally the bias is introduced by applying a DC (Direct Current) voltage $V_0 \in [-V_{\pi,DC}, V_{\pi,DC}]$ to the "DC electrode" of the M-Z. When a bias is applied, the scaling of the input amplitude is important, and the voltage driving the RF (Radio Frequency) electrode of the M-Z is taken to be $V=\gamma m(t)u(t) \in [-\gamma V_{\pi,RF}, \gamma V_{\pi,RF}]$. The input signal therefore has the form:

$$A_{in}(t) = \sin\left[\frac{\pi}{2}\left(\frac{\gamma m(t)u(t)}{V_{\pi,RF}} + \frac{V_0}{V_{\pi,DC}}\right)\right]. \quad (4)$$

In our case, we measured $V_{\pi,DC}$=7.54V and $V_{\pi,RF}$=7.27V. The AWG followed by RF amplifiers provides a maximum voltage amplitude $V$ of 3.27V on both RF electrodes of the M-Z, corresponding to the possibility to tune $\gamma$ in the range of [0,0.45]. All the results presented here were obtained with a maximum power at the output of the M-Z (i.e. at the entrance of the cavity) of 2.11mW for unbiased masked input signal (respectively 5mW for biased masked input signal). This value was chosen to provide sufficient signal-to-noise ratio on the readout photodiode, while avoiding Brillouin backscattering issues. However, we experimentally showed that we can decrease the power at the output of the M-Z to as low as 0.57mW for unbiased masked input signal (respectively 1.35mW for biased masked input signal) without affecting performance. Below this level, the signal-to-noise ratio on the photodiode was too small.

The reservoir itself consists of a ~230m long passive fiber cavity (made of Single Mode Fiber SMF-28e), corresponding to a roundtrip time of $T$=1.13209µs. With the desynchronization parameter $k$=1, and using $N$=50 internal variables (which we use for most tasks), we therefore have $T''$=1.10989µs and the temporal length of each internal variable is equal to $\theta$=22.2ns. The output refresh rate $1/T''$~0.9MHz can be seen as the processing speed of our RC. It is quite low due to the use of rather slow electronics (AWG and photodiode) and a correspondingly long fiber cavity. However, this is not a fundamental limit since refresh rate can in principle easily be increased by using a smaller cavity and faster electronics.

The input signal $A_{in}(t)$ is injected into the passive fiber cavity using a 90/10 coupler, corresponding to $\beta$=(0.1)$^{1/2}$=0.316. A tunable optical attenuator is used to adjust the feedback gain in the range 0<$\alpha$<0.806.

In order to stabilize the fiber cavity, 10% of the laser light intensity is coupled into it in the counterpropagating direction using a 90/10 coupler (see Fig. 1), and its intensity is used as a control signal processed by a PID (Proportional-Integral-Derivative) regulator which drives a piezo-electric fiber stretcher inserted in the cavity. In this way the phase detuning $\Delta\varphi$ of the cavity can be precisely controlled. A 90/10 coupler is finally used to send 10% of the optical intracavity power to the readout photodiode. A digital oscilloscope records the photodiode signal. The oscilloscope record is used in a digital post-processing stage realized by a computer to compute the readout weights $W_i$, to produce the output $y(n)$, and to estimate the performance on specific tasks. Note that no time averaging is used in the oscilloscope recording, contrary to some earlier experiments.

The results of the experiment were carefully compared with discrete time simulations. Further details on the experimental setup and simulations are provided in the appendix.

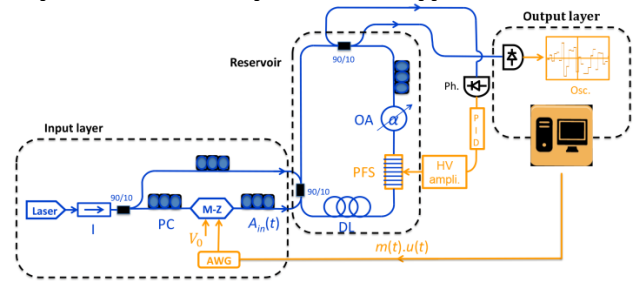

**Fig. 1.** Experimental setup. Blue lines (dark grey) correspond to fiber optic and orange lines (light grey) to electrical connections. Laser: long coherence telecom wavelength laser; I: Isolator, PC: Polarization Controller; M-Z: amplitude modulator (lithium niobate Mach-Zehnder interferometer in push-pull configuration); OA: Optical Attenuator; PFS: Piezo-electric Fiber Stretcher; AWG: Arbitrary Waveform Generator; DL: Delay Loop; HV ampli.: High Voltage Amplifier; PID: Proportional-Integral-Derivative regulator; Ph.: Photodiode; Osc.: Oscilloscope. The PC inside the cavity is used to control the Jones matrix of the cavity, and the PCs before the cavity ensure that $A(t)$ and the counterpropagating control signal are on the two different polarization eigenmodes of the cavity.

## 4. Results

Here we present the results we obtained both experimentally and through simulations for different benchmark tasks widely used in the reservoir computing community. Except where indicated, we used $N$=50 internal variables and a ridge parameter of $10^{-4}$. As there are many parameters to scan ($\alpha$, $\Delta\varphi$, $V_0/V_{\pi,DC}$, $\gamma$ and the ridge parameter), the reliability of our simulations allowed us to scan experimentally only the regions in parameter space that provide good performances. The experimental results are then compared with the simulation results within these regions. The experimentally reachable parameter ranges with our setup are the following: $\alpha \in [0.366, 0.806]$, $\Delta\varphi \in [-\pi,\pi]$rad, $V_0/V_{\pi,DC} \in [-1,1]$, and $\gamma \in [0, 0.45]$. When $V_0/V_{\pi,DC}$ =0, $\gamma$ was always set to its maximum experimental reachable value.

For each choice of these parameters, we repeated the experiment ten times with ten different input series, averaging the performance obtained, except for the speech recognition task. These performances are then compared with the best published results obtained up to now with different reservoir computer architectures [7,12-17,21,22,24,29,31,32].

In the following subsections, we provide a small description for all of these tasks. For a more in-depth description, see [13,15].

### A. Memory capacities evaluation

In this basic task we test the ability of the reservoir to recall simple linear or nonlinear functions $P$ of previous inputs $u(n-l)$. Three different memory capacities (MC) are considered [33,34], i.e. the linear, quadratic and cross memory capacities. The memory function $MF(l)$ is computed as $MF$=1-NMSE$\in[0,1]$ (NMSE: Normalized Mean Square Error) and is obtained by computing the correlation between $P$ and the reservoir states, as is done in [33]. The total memory is the sum of the three different memory capacities. Theoretically it cannot exceed the number of internal variables [33]. Table 1 reports the best values obtained for each memory capacity, as well as for the total memory capacity, compared with three other architectures [13,15,17]. The results were obtained using 10 datasets of 2200 input samples randomly drawn from a uniform distribution in the interval [-1,1] (except for the total memory in simulation, in which case 10 datasets of 60000 input samples were used, to lower statistical noise, given that the total memory in our particular reservoir architecture is very close to the maximum).

**Table 1. Memory capacities evaluation**

| | Linear memory | Quadratic memory | Cross memory | Total memory |
|---|---|---|---|---|
| MF | $LMF(l)$: $P(u(n-l))=u(n-l)$ | $QMF(l)$: $P(u(n-l))=$ $3u^2(n-l)-1$ | $XMF(l,l')$: $P(u(n-l), u(n-l'))=$ $u(n-l)u(n-l')$ | |
| MC | $LC = \sum_{l=0}^{l=l_{max}} LMF(l)$ | $QC = \sum_{l=0}^{l=l_{max}} QMF(l)$ | $XC = \sum_{l=0}^{l=l_{max}} \sum_{l'=l+1}^{l'=l'_{max}} XMF(l,l')$ | $C=$ $LC+QC$ $+XC$ |
| Sim. | $LC$=22.61±0.07 | $QC$=13.25±0.23 | $XC$=33.90±0.82 | $C=$ 49.99±0.13 |
| Exp. | $LC$=21.14±0.34 | $QC$=12.07±0.10 | $XC$=30.20±0.46 | $C=$ 48.37±0.47 |
| [13] | $LC$=31.9 | $QC$=4 | $XC$=27.3 | $C$=48.6 |
| [15] | $LC$=20.84 | $QC$=4.16 | $XC$=4.71 | $C$=25.20 |
| [17] | $LC$=36.8 | $QC$=2.23 | $XC$=14.32 | $C$=37.05 |

MF: Memory Function; MC: Memory Capacity; Sim.: Simulation results of our architecture; Exp.: Experimental results of our architecture; LMF: Linear Memory Function; LC: Linear Capacity; QMF: Quadratic Memory Function; QC: Quadratic Capacity; XMF: cross Memory Function; XC: cross Capacity; C: (total) Capacity. The conditions in which these results were obtained both in simulation and experiment are: LC was obtained for $\alpha$=0.806 and by scanning $\Delta\varphi$ in the range [0,1.42]rad, $V_0/V_{\pi,DC}$ in the range [-0.6,-0.53] and $\gamma$ in the range [0.33,0.45] (without any precompensation of the masked input signal); QC was obtained for $\alpha$=0.806, $\Delta\varphi$=0 and by scanning $V_0/V_{\pi,DC}$ in the range [-0.22,0.85] and $\gamma$ in the range [0.3,0.45] (without any precompensation of the masked input signal); XC and C were obtained for $\alpha$=0.806, $\Delta\varphi$=0 and $V_0/V_{\pi,DC}$=0 (with precompensation of the masked input signal).

Note that all the memory capacities results obtained experimentally for the three architectures [13,15,17] selected for comparison and reported in table 1, are taken from [17]. The reasons are that memory capacity results for the optoelectronic architecture [13] were first reported in [17], and that the memories capacities results for the SOA (Semiconductor Optical Amplifier) based architecture published in [15] contain mistakes that were corrected in [17].

The linear memory we obtained is comparable to the one obtained for the SOA-based RC [15], but lower than for the optoelectronic [13] and Saturable Absorber (SA) based [17] RCs. The quadratic memory is about 3 times larger than the values reported for the optoelectronic [13] and SOA-based [15] RCs, and more than 5 times larger than for the SA based RC [17]. The cross memory is slightly larger than the one obtained for the optoelectronic architecture [13], but more than 6 times larger than with the SOA-based RC [15], and 2 times larger compared than the SA-based RC [17]. Finally, the total memory capacity is comparable to the one obtained for the optoelectronic RC [13], and larger than the ones obtained for the SOA [15] and SA [17] based RC.

To sum up, for most of the memory capacities evaluations, our reservoir architecture exhibits comparable or better performances than the three experimental RC setups [13,15,17] taken for comparison, sometimes several times bigger. The fact that the experimental total capacity $C$ is very close to the maximum value (50) indicates that very little noise is affecting the experiment.

Note that the simulation model is in good agreement with the experimental results. As there is no noise in the simulation, the total memory capacity in simulation is found to be extremely close to the maximum value of 50, in agreement with theoretical expectations.

### B. NARMA10

The NARMA10 (Nonlinear Auto-Regressive Moving Average) task is one of the most-used benchmark tasks in the field of reservoir computing [1,7,11-13,32]. The aim is to reproduce the behavior of a nonlinear, tenth-order system with random input $u(n)$ drawn from a uniform distribution over the interval [0,0.5]. The target output is the following:

$$y^*(n+1) = 0.3y^*(n) + 0.05y^*(n)\left[\sum_{i=0}^{9} y^*(n-i)\right] + 1.5u(n-9)u(n) + 0.1. \quad (5)$$

We trained the reservoir over 1000 steps (1000 values of $u(n)$), and tested its performance for the subsequent 1000 steps. The standard deviation was evaluated by repeating this procedure 10 times. The performances obtained were measured using the Normalized Mean Square Error (NMSE), defined by:

$$NMSE = \frac{\sum_n [y^*(n) - y(n)]^2}{\sum_n \left\{y^*(n) - \frac{\sum_n [y^*(n)]}{n}\right\}^2}. \quad (6)$$

For this task, all the results below were obtained with a precompensated masked input signal and no bias $V_0/V_{\pi,DC}$=0. Using $N$=50 internal variables, we obtained a NMSE of 0.104±0.02 for the simulation and an experimental NMSE of 0.107±0.012. These performances were obtained with $\alpha$=0.806 and $\Delta\varphi$ was scanned in the ranges [0,1.81]rad and [2.61,3.67]rad. This surpasses the best result reported in experiment until now with $N$=50. For instance in

[13] a NMSE=0.168±0.015 was obtained with the same number of variables. The value NMSE=0.16 in fact corresponds to the best that can be obtained with a linear shift register [12].

Our simulation model predicts very well our experimental results. Simulations also showed that our results are strongly dependent on the oscilloscope resolution: by increasing the acquisition resolution from 8 to 14 bits in our simulation model (thus decreasing the quantization noise), we reach a NMSE of 0.062±0.008, which is a very good result. This simulation result was obtained with $\alpha$=0.806 and $\Delta\varphi$ scanned in the range [0,π]rad. The optimum was reached for $\Delta\varphi$=0.38rad. This is in the range already scanned for the results mentioned above with a resolution acquisition of 8 bits, which implies that the increase in performance is only due to the increase of the oscilloscope resolution. This result is, to our knowledge, better than any result published up to now with $N$=50 internal variables (note that using 50 variables, the algorithm reported in [7] obtained NMSE=0.152±0.0138). We attribute these very good performances to our architecture in which the reservoir is linear and the readout quadratic. Indeed the nonlinearities exhibited by the NARMA10 equation are mainly quadratic.

The performance for the NARMA10 task strongly depends on the number of internal variables. Upon increasing the number of internal variables to $N$=300, we obtained NMSE=0.0484±0.0095 in experiment, and NMSE=0.0463±0.0142 in simulation. These results were obtained with $\alpha$=0.806 and $\Delta\varphi$ scanned in the range [0.26,2.35]rad.

Finally it should be noted that using the simple algorithm given by (2) and (3) (that is without the experimental constraints set out in the appendix and at the end of the first paragraph of this section) we obtained NMSE=0.0106±0.0030 with $N$=400 internal variables. This performance was obtained by scanning $\alpha$ in the range [0.7,0.99] and $\Delta\varphi$ in the range [0,π]rad. Using the same number of internal variables, a simulation result of NMSE=0.022 is reported in [12]. A NMSE=0.018 is also reported in [32] using $N$=520 internal variables.

### C. Nonlinear channel equalization

In this task, first used in the context of RC in [3], the goal is to recover an input symbol sequence $d(n)$ from the signal received at the output $u(n)$ of a standardized nonlinear multipath RF channel, defined as follows:

$$q(n) = 0.08d(n+2) - 0.12d(n+1) + d(n) + 0.18d(n-1) - 0.1d(n-2) + 0.091d(n-3) - 0.05d(n-4) + 0.04d(n-5) + 0.03d(n-6) + 0.01d(n-7), \quad (7)$$

$$u(n) = q(n) + 0.036q^2(n) - 0.011q^3(n) + v(n). \quad (8)$$

Here $v(n)$ is a Gaussian noise with zero mean, adjusted to yield Signal-to-Noise Ratios (SNR) ranging from 12 to 32 dB. The symbols $d(t)$ are randomly chosen between four values {-3,-1,1,3}. The performance is evaluated in terms of Symbol Error Rate (SER), which is the fraction of misclassified symbols.

For this task, we used 50 internal variables, 3000 training samples and 6000 test samples. The standard deviation reported in Fig. 2 was evaluated by repeating this procedure 10 times. The results obtained are reported in Fig. 2 and compared to the ones obtained in [13,17]:

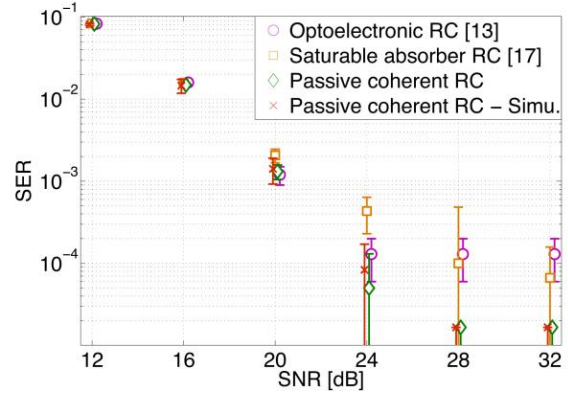

**Fig. 2.** Experimental and simulation results for the nonlinear channel equalization task. Horizontal axis signal-to-noise ratio of the nonlinear channel, vertical axis symbol error rate, i.e. number of misidentified symbols. These results were obtained with $\Delta\varphi$=0rad, $\gamma$=0.45 and without any precompensation of the masked input signal. To get the best performances for each SNR, $\alpha$ was scanned in the range [0.45,0.55], $V_0/V_{\pi,DC}$ in the range [-0.5,0.5] and the ridge parameter from $10^{-6}$ to $10^{-4}$. Note that at 28 and 32 dB SNR, the SER is zero for the Passive coherent RC, both in simulation and experiment.

Again, we can see that our passive coherent RC architecture outperforms the previous architectures, using the same test conditions. In particular for a SNR of 28dB and 32dB we obtained both in simulation and experimentally, a SER of 0%, which means that all the 60000 symbols were correctly identified. Such very good performance was never attained in previous experiments.

### D. Isolated spoken digits recognition

The goal of this task is to classify 10 digits pronounced 10 times by 5 different female speakers, first without noise and then with babble noise added to the sound records to reach a Signal-to-Noise Ratio (SNR) of 3dB, following the procedure explained in [24]. The data was taken from the NIST TI-46 Corpus [35] (National Institute of Standards and Technology Texas Instrument-46 corpus), and the recording was pre-processed according to the Lyon cochlear ear model [36]. Ten output classifiers were trained, one for each digit. Each classifier was trained to output « +1 » if a specific digit was sent to the reservoir, « -1 » otherwise. A winner-takes-all approach between the classifiers was then used to select the actual response of the reservoir.

For a better estimation of the performance, a cross validation procedure was applied over 5 subsets of 100 words, chosen randomly. The reservoir was trained on 4 of the subsets, and then tested on the fifth one. This procedure was repeated 5 times, so that each subset is used once as test subset. The performance values, evaluated in terms of Word Error Rate (WER), which is the fraction of misclassified digits, are reported in table 2.

**Table 2. Experimental and simulation results for the isolated spoken digit recognition task**

|  | No noise on the signal, with 200 internal variables | SNR=3dB, with 500 internal variables |
|---|---|---|
| **Simulation** | WER=0% | WER=0.6(±0.9)% |
| **Experiment** | WER=0% | WER=0.8(±0.8)% |

These results were obtained with precompensation of the masked input signal, no bias and $\alpha$=0.806. The ridge parameter was scanned from $10^{-6}$ to $10^{-4}$ and $\Delta\varphi$ was scanned in the range [-1.62,0.19]rad.

Again, we can see that our simulation results are very close to the results obtained experimentally. With no noise on the input signal, we obtained, using 200 internal variables, an experimental WER of 0%, which means that all the 500 digits were well classified. A WER of 0% obtained in simulation is also reported in [22] using 400 internal variables. Our result surpasses the one obtained with the optoelectronic RC [13] (WER of 0.4%) with $N$=200 internal variables. A WER of 3±(1.2)% is also reported in [15] using $N$=200.

Experimental results on this task are also reported in [12,14,16,21,22,31]. The reported WERs are 0.2% [12], 0.04(±0.017)% [14], 0.014(+0.051/-0.014)% [16,22], 0.6(±0.2)% [31], 0.05% [21]. In these works, $N$=400 in [12,14,21], $N$=388 in [16,22], $N$=150 in [31] and for the cross validation procedure, the reservoir was trained on 475 spoken digits and tested on 25 digits. Concerning speed, it should be noted that the setup presented in [16] carried out this task with a refresh rate slightly more than 14 times faster than in the present experiment, while our setup is almost 19 times faster than [14], 7.2 10$^5$ times faster than [12] and more than 200 times faster than in [31].

This task can be made more complicated by adding background babble noise. With a SNR of 3dB, we obtained experimentally WER=0.8(±0.8)% with $N$=500. Unfortunately, we cannot compare this performance because it has never been tested experimentally until now. In simulation we obtain a WER=0.6(±0.9)% with $N$=500. This can be compared to the simulation results of WER=4.5% [24] using $N$=81 and of WER=1% [29] using $N$=500.

On this task, our algorithm (i.e. the simple algorithm given by (2) and (3), without the experimental constraints set out in the appendix and at the end of the first paragraph of this section) reaches WER=0 with $N$=90 for noiseless signal, and WER=0 with $N$=350 for noisy signal. These later results were obtained with precompensation of the masked input signal and without any bias. $α$ was scanned from 0.8 to 0.99, $Δφ$ was scanned in the region [0,π]rad and the ridge parameter was scanned from 10$^{-6}$ to 10$^{-4}$.

## 5. Conclusions

In the present work we have demonstrated a photonic implementation of a passive linear fiber reservoir computer working with coherent light for analog signal processing. This experiment presents many qualities that were either absent or not simultaneously present in previous works: we can perform analog optical signal processing, the easy tunability of each key parameters achieves the best operating point for each task, the system is able to reach a strikingly weak noise floor thanks to the absence of active elements in the reservoir itself, richer dynamics is provided by operating in coherent light, and finally high power efficiency is yielded as a result of the passive nature and simplicity of the setup. Note that at this stage we have only obtained low optical power consumption for the reservoir itself, and have not tried to minimize the overall power consumption, including all control electronics.

The main challenge was to stabilize the system, as our reservoir is a long interferometer made of an optical fibered cavity of approximately 230m. In the future, faster electronics (photodiode, AWG and oscilloscope) will enable smaller, possibly even integrated, cavities, and hence much simpler stabilization. In addition, our expertise in interferometric stabilization will allow us in the future to study photonic reservoir computers in which the reservoir states are processed in parallel, rather than sequentially, providing significant further speedup.

On the different tasks we tested, the present experiment has error rate as low or lower than previous experiments. For instance the NARMA10 task was up to now considered as a big challenge, and no experiment with performance exceeding a linear shift register had been reported before the present work.

Quite remarkably, through its conceptual simplicity, our experimental approach has also contributed to the theory of reservoir computing. Indeed, the discrete time equations (2) and (3) constitute a very simple high performance reservoir computer algorithm. In particular, it combines three advantages: (i) the simple interconnection matrix first introduced in [7], (ii) the simplicity of a linear reservoir associated with a nonlinear output layer and (iii) very good performances on benchmark tasks even when small number of internal variables are used. Our algorithm thus seems very computationally efficient and could also find applications in numerical implementations of reservoir computers. This should however be confirmed by a more detailed comparison with other RC algorithms.

Given the good performances and conceptual simplicity of the experiment reported here, one can expect that this architecture constitutes an important milestone in the future progress on photonic reservoir computing.

## Appendix A

### 1. Numerical simulations

Our simulation model is based on the discrete time equations (2) and (3), with the nonlinear pre-processing (4) taken into account when relevant. In addition, the simulation includes the effect of the most important features of the components of our experimental setup: the 14 bits of resolution of the AWG, the 8 bits of resolution of the oscilloscope, the saturation transfer function of the RF amplifiers used to amplify the signal sent to the MZ from the AWG, the low cutoff frequency of these RF amplifiers (30kHz-12GHz bandwidth), the low cutoff frequency of the photodiode used to read the $|x_i(n)|^2$ (30kHz-1GHz bandwidth).

### 2. Interferometric stabilization and phase detuning setting

Our reservoir is a fiber optics cavity with a roundtrip time $T$=1.13209μs corresponding to approximately 230m of fiber. This cavity needs thus to be phase stabilized. This is done using a piezoelectric fiber stretcher that compensates phase shifts introduced in the cavity by thermal, vibrational and phonic noise. To this end, a non-modulated optical signal, the control signal, is sent in the cavity in the opposite direction compared to the reservoir states signal coded in $A(t)$, as shown in Fig. 1. This signal is collected by a photodiode which drives a digital PID regulator. The output signal of the PID regulator is then amplified in order to control the piezoelectric fiber stretcher.

The polarization controller inside the cavity (see Fig. 1) is used to tune the Jones matrix of the cavity (the polarization transfer matrix of the cavity), and thereby tune "the phase offset" between the two polarization eigenmodes of the cavity. The reservoir states signal and the counterpropagating control signal are injected on the two different polarization eigenmodes of the cavity. The PID stabilizes the cavity on one of the slopes of the control signal resonance. By changing the Jones matrix of the cavity, the phase distance between the resonances of the reservoir states signal and the control signal can be arbitrarily adjusted, and thus the detuning $Δφ$ appearing in eq. (1) can be adjusted to an arbitrary value.

This is illustrated in Fig. 3 which shows the measured cavity transfer function for two non-modulated input signals coupled in counterpropagating directions in the cavity, on the two different polarization eigenstates.

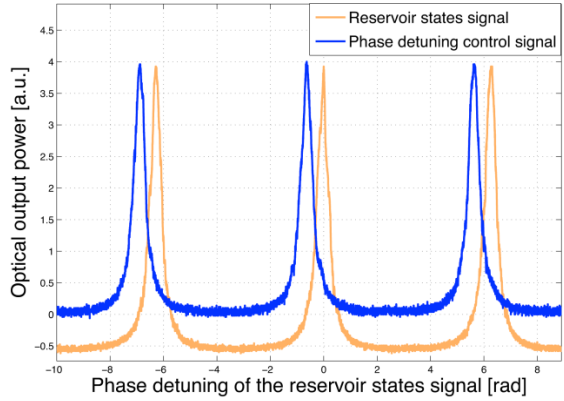

**Fig. 3.** Resonances of the cavity. Two CW signals are coupled in counterpropagating directions in the cavity, on the two polarization eigenmodes of the cavity. A voltage ramp is applied to the piezoelectric fiber stretcher in order to scan the cavity phase, and the output power is recorded. The optical attenuator was set at its maximum transparency.

In order to minimize the phonic, vibrational and thermal noise, we isolated every cavity component in several boxes made of 1 cm to 2 cm thick aluminum plates with stone wool inside, as shown in Fig. 4 for the delay loop:

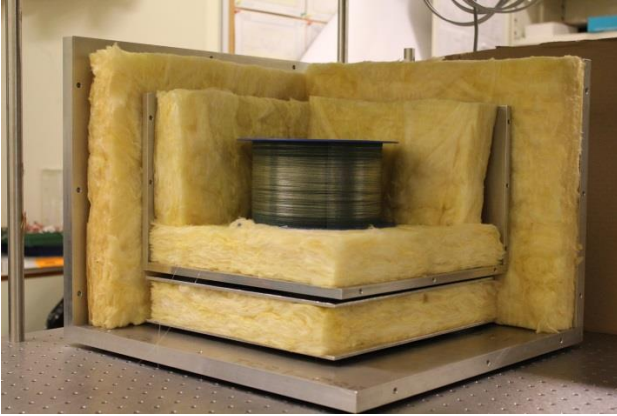

**Fig. 4.** Isolation system of the delay loop, open. It consists of two aluminum boxes: a small box made of 1cm thick plates inside a bigger box made of 2cm thick plates. The inner plates of both boxes are covered with stone wool. Each box is isolated from the base on which it rests by sorbothane sheets that absorb vibrations.

All the boxes fit together so that the isolation system of the whole cavity is almost airtight. These boxes were designed to isolate the cavity from the phase noise frequencies that the PID regulator is not able to follow, as its output refresh rate of 8.3 kHz is limited by the resonance frequency of the piezoelectric fiber stretcher. Moreover, the whole experiment is placed on an optical table maintained on air cushions.

Note that in this first implementation, the comparatively long fiber cavity (230m) was designed so as to be able to use rather slow electronics (AWG and photodiode). In future work we intend to use faster electronics, and hence a smaller cavity. A smaller cavity will be much easier to stabilize and isolate from sources of phase noise. An implementation on a chip, as in [29], will not require any stabilization (except for thermal control of the whole chip).

## 3. Experimental setup components

Table 3 presents the relevant technical characteristics of the main components used in our hardware setup:

**Table 3. Technical description of the most important hardware components**

| Equipment | Technical description |
|---|---|
| Laser | Koheras adjustik KOH1995 laser:<br>- Wavelength: 1550nm<br>- FWHM (Full Width at Half Maximum): 1kHz<br>- Maximum output power: 100mW |
| Reservoir | Optical fibered cavity made with 2 90/10 couplers and 1 optical attenuator:<br>- Maximum feedback gain in electromagnetic field amplitude: 0.806<br>- Minimum FWHM: 60.3kHz<br>- Maximum Finesse: 14.65 |
| Arbitrary Waveform Generator | Agilent AWG M8190A:<br>- 14 bits resolution<br>- Sample rate from 125 MSamples/s to 8 GSamples/s<br>- DC to 5 GHz analog bandwidth<br>- Up to 128MSamples arbitrary waveform memory per channel |
| Acquisition system : oscilloscope | Agilent DSA91204A:<br>- 12 GHz bandwidth<br>- up to 40 GSamples/s on each of 4 analog channels<br>- 1 GSamples memory per channel |
| PID regulator | Homemade PID regulator programmed using mbed NXP LPC1768 development board:<br>- Processor: 32-bits ARM Cortex-M3 core running at 96MHz<br>- ADC (Analog-to-Digital Converter): 12 bits resolution<br>- DAC (Digital-to-Analog Converter): 10 bits resolution<br>- Output refresh rate: 8.3 kHz |
| Mach-Zehnder Interferometer | Lucent electro-optic modulator:<br>- Model: X2624C<br>- 20 GHz analog bandwidth |
| Reservoir states readout photodiode | New focus low noise photoreciever, model 1611 IR 1GHz:<br>- Bandwidth (3dB): 30kHz-1GHz<br>- Risetime: 400ps<br>- Current gain: 700V/A<br>- Responsivity at 1550nm: 1.04 A/W |

## Funding Information

We acknowledge financial support by Interuniversity Attraction Poles program of the Belgian Science Policy Office under grant IAP P7-35 "photonics@be", by the Fonds pour la formation à la Recherche dans l'Industrie et dans l'Agriculture (FRIA), by the Fonds de la Recherche Scientifique FRS-FNRS (FRFC grant T.0092.14), by the Action de Recherche Concertée through grant AUWB-2012-12/17-ULB9, and by the ERC through the Naresco starting grant (ref. 239599).

## References

1. H. Jaeger, "The "echo state" approach to analysing and training recurrent neural networks," Technical Report GMD Report **148**, German National Research Center for Information Technology (2001).
2. W. Maass, T. Natschläger, and H. Markram, "Real-time computing without stable states: a new framework for neural computation based on perturbations," Neural. Comput. **14**, 2531-2560 (2002).
3. H. Jaeger, and H. Haas, "Harnessing nonlinearity: predicting chaotic systems and saving energy in wireless communication," Science **304**, 78-80 (2004).
4. F. Triefenbach, A. Jalal, B. Schrauwen, and J.-P. Martens, "Phoneme recognition with large hierarchical reservoirs," Adv. Neural Inf. Proces. Syst. **23**, 2307-2315 (2010).
5. L. Boccato, A. Lopes, R. Attux, and F. J. Von Zuben, "An Echo State Network Architecture Based on Volterra Filtering and PCA with

Application to the Channel Equalization Problem," International Joint Conference on Neural Networks, 580-587 (2011).
6. L. Boccato, A. Lopes, R. Attux, and F. J. Von Zuben, "An extended echo state network using Volterra filtering and principal component analysis," Neural Netw. **32**, 292-302 (2012).
7. A. Rodan, and P. Tino, "Simple deterministically constructed recurrent neural networks," Intelligent Data Engineering and Automated Learning (IDEAL), 267-274 (2010).
8. P. Buteneers, D. Verstraeten, P. Van Mierlo, T. Wyckhuys, D. Stroobandt, R. Raedt, H. Hallez, and B. Schrauwen, "Automatic detection of epileptic seizures on the intra-cranial electroencephalogram of rats using reservoir computing," Artif. Intell. Med. **53**, 215-223 (2011).
9. E. A. Antonelo, B. Schrauwen, and D. Stroobandt, "Event detection and localization for small mobile robots using reservoir computing," Neural Netw. **21**, 862-871 (2008).
10. M. Lukoševičius, H. Jaeger, and B. Schrauwen, "Reservoir Computing Trends," KI - Künstliche Intelligenz **26**, 365-371 (2012).
11. M. Lukoševičius, and H. Jaeger, "Reservoir computing approaches to recurrent neural network training," Computer Science Review **3**, 127-149 (2009).
12. L. Appeltant, M. C. Soriano, G. Van der Sande, J. Danckaert, S. Massar, J. Dambre, B. Schrauwen, C. R. Mirasso, and I. Fischer, "Information processing using a single dynamical node as complex system," Nat. Commun. **2**, 468 (2011).
13. Y. Paquot, F. Duport, A. Smerieri, J. Dambre, B. Schrauwen, M. Haelterman, and S. Massar, "Optoelectronic reservoir computing," Sci. Rep. **2**, 287 (2012).
14. L. Larger, M. C. Soriano, D. Brunner, L. Appeltant, J. M. Gutierrez, L. Pesquera, C. R. Mirasso, and I. Fischer, "Photonic information processing beyond Turing: an optoelectronic implementation of reservoir computing," Opt. Express **20**, 3241-3249 (2012).
15. F. Duport, B. Schneider, A. Smerieri, M. Haelterman, and S. Massar, "All-optical reservoir computing," Opt. Express **20**, 22783-22795 (2012).
16. D. Brunner, M. C. Soriano, C. R. Mirasso, and I. Fischer, "Parallel photonic information processing at gigabyte per second data rates using transient states," Nat. Commun. **4**, 1364 (2013).
17. A. Dejonckheere, F. Duport, A. Smerieri, L. Fang, J.-L. Oudar, M. Haelterman, and S. Massar, "All-optical reservoir computer based on saturation of absorption," Opt. Express **22**, 10868-10881 (2014).
18. D. Brunner, M. C. Soriano, and I. Fischer, "High-speed optical vector and matrix operations using a semiconductor laser," IEEE Photon. Technol. Lett. **25**, 1680-1683 (2013).
19. M. C. Soriano, S. Ortín, D. Brunner, L. Larger, C. R. Mirasso, I. Fischer, and L. Pesquera, "Optoelectronic reservoir computing: tackling noise-induced performance degradation," *Opt. express* **21**, 12-20 (2013).
20. M. Hermans, M. Soriano, J. Dambre, P. Bienstman, and I. Fischer, "Photonic Delay Systems as Machine Learning Implementations," arXiv: 1501.02592v1 [Neural and Evolutionary Computing] (2015).
21. M. C. Soriano, S. Ortin, L. Keuninckx, L. Appeltant, J. Danckaert, L. Pesquera, and G. Van Der Sande, "Delay-based Reservoir Computing: noise effects in a combined analog and digital implementation," IEEE Trans. Neural Netw. Learn. Syst. **26**, 388-393 (2015).
22. K. Hicke, M. A. Escalona-Moran, D. Brunner, M. C. Soriano, I. Fischer, and C. R. Mirasso, "Information processing using transient dynamics of semiconductor lasers subject to delayed feedback," IEEE J. Sel. Top. Quantum Electron. **19**, 1501610 (2013).
23. R. M. Nguimdo, G. Verschaffelt, J. Danckaert, and G. Van der Sande, "Fast photonic information processing using semiconductor lasers with delayed optical feedback: Role of phase dynamics," Opt. Express **22**, 8672-8686 (2014).
24. K. Vandoorne, J. Dambre, D. Verstraeten, B. Schrauwen, and P. Bienstman, "Parallel reservoir computing using optical amplifiers," IEEE Trans. Neural Netw. **22**, 1469-1481 (2011).
25. K. Vandoorne, W. Dierckx, B. Schrauwen, D. Verstraeten, R. Baets, P. Bienstman, and J. Van Campenhout, "Toward optical signal processing using photonic Reservoir Computing," Opt. Express **16**, 11182-11192 (2008).
26. C. Mesaritakis, V. Papataxiarhis, and D. Syvridis, "Micro ring resonators as building blocks for an all-optical high-speed reservoir-computing bit-pattern-recognition system," JOSA B **30**, 3048-3055 (2013).
27. H. Zhang, X. Feng, B. Li, Y. Wang, K. Cui, F. Liu, W. Dou, and Y. Huang, "Integrated photonic reservoir computing based on hierarchical time-multiplexing structure," Opt. Express **22**, 31356-31370 (2014).
28. M. A. A. Fiers, T. Van Varenbergh, F. Wyffels, D. Verstraeten, B. Schrauwen, J. Dambre, and P. Bienstman, "Nanophotonic Reservoir Computing With Photonic Crystal Cavities to Generate Periodic Patterns," IEEE Trans. Neural Netw. Learn. Syst. **25**, 344-355 (2014).
29. K. Vandoorne, P. Mechet, T. Van Vaerenbergh, M. Fiers, G. Morthier, D. Verstraeten, B. Schrauwen, J. Dambre, and P. Bienstman, "Experimental demonstration of reservoir computing on a silicon photonics chip," Nat. Commun **5**, 3541 (2014).
30. A. Smerieri, F. Duport, Y. Paquot, B. Schrauwen, H. Haelterman, and S. Massar, "Analog readout for optical reservoir computers," Adv. Neural Inf. Proces. Syst. **25**, 953-961 (2012).
31. R. Martinenghi, S. Rybalko, M. Jacquot, Y. K. Chembo, and L. Larger, "Photonic nonlinear transient computing with multiple-delay wavelength dynamics," Phys. Rev. Lett. **108**, 244101 (2012).
32. L. Appeltant, G. Van der Sande, J. Danckaert, and I. Fischer, "Constructing optimized binary masks for reservoir computing with delay systems," Sci. Rep. **4**, 3629 (2014).
33. J. Dambre, D. Verstraeten, B. Schrauwen, and S. Massar, "Information processing capacity of dynamical systems," Sci. Rep. **2**, 514 (2012).
34. H. Jaeger, "Short Term Memory in Echo State Networks," Technical Report GMD Report **152**, German National Research Center for Information Technology (2002).
35. Texas Instruments-Developed 46-Word Speaker-Dependent Isolated Word Corpus (TI46) September 1991, NIST Speech Disc 7-1.1 (1 disc).
36. R. Lyon, "A computational model of filtering, detection, and compression in the cochlea," IEEE International Conference on Acoustics, Speech, and Signal Processing (ICASSP) **7**, 1282-1285 (1982).